\documentstyle[longtable]{aipproc}

\input psfig

\def\gtorder{\mathrel{\raise.3ex\hbox{$>$}\mkern-14mu
             \lower0.6ex\hbox{$\sim$}}}
\def\ltorder{\mathrel{\raise.3ex\hbox{$<$}\mkern-14mu
             \lower0.6ex\hbox{$\sim$}}}

\begin{document}

\title{Results from the CASTLES Survey of Gravitational Lenses}
 
\author{
 C.S. Kochanek$^*$,
 E.E. Falco$^*$,
 C.D. Impey$^\dagger$,
 J. Leh\'ar$^*$,
 B.A. McLeod$^*$ \&
 H.-W. Rix$^\dagger$
 }
\address{
        $^*$Center for Astrophysics, 60 Garden St., Cambridge, MA 02138 \\
$^{\dagger}$Steward Observatory, Univ. of Arizona, Tucson, AZ 85721 }

%\lefthead{LEFT head}
%\righthead{RIGHT head}
\maketitle

\begin{abstract}
We show that most gravitational lenses lie on the passively evolving fundamental
plane for early-type galaxies. For burst star formation models (1~Gyr of star
formation, then quiescence) in low $\Omega_0$ cosmologies, the stellar 
populations of the lens galaxies must have formed at $z_f \gtorder 2$.
Typical lens galaxies contain modest amounts of patchy extinction, with a
median differential extinction for the optical (radio) selected lenses of
$\Delta E(B-V) = 0.04$ ($0.07$) mag.  The dust can be used to determine both 
extinction laws and lens redshifts.
For example, the $z_l=0.96$ elliptical lens in MG~0414+0534
has an $R_V=1.7\pm0.1$ mean extinction law.  Arc and ring images
of the quasar and AGN source host galaxies are commonly seen in
NICMOS H band observations.  The hosts are typically blue, $L \ltorder L_*$
galaxies.
\end{abstract}

\section*{1. Introduction}

In the last few years the number of known gravitational lenses has exploded to a 
total of over 40 systems\footnote{see http://cfa-www.harvard.edu/castles.}.
With such a large statistical sample the lenses become excellent tools for
studying the structure and evolution of the lens galaxies, the luminosity
function of lens galaxies, dust in the lens galaxies, and cosmology.  Many 
of these applications, particularly the evolution and luminosity function
studies, depend on possessing accurate surface photometry of the lens 
galaxies. When we attempted the first survey of galaxy evolution and 
structure using lenses (Keeton et al. 1998), we discovered that the 
accumulated photometric
data were inadequate to the task.  Individual groups had observed
individual lenses in a remarkable array of filters, frequently using
short snapshots which were adequate to verify lens candidates
but inadequate for detailed surface photometry of the lens
galaxies.  

The goal of the CfA/Arizona Space Telescope Lens Survey (CASTLES) is
to remove these limitations and to fully enable the use of lenses as
precision tools in astronomy.  To date we have obtained NICMOS H
images of 37 lenses and binary quasars, with complementary WFPC2 
V and I images scheduled for 1999.  Further observations proposed for 
Cycle 8 would expand
the database to 60 lenses and binary quasars.  The data are
homogeneously reduced with typical astrometric accuracies of 3 mas
or better (checked on VLBI lenses), and initially fit with
a series of standard photometric and gravitational lens models
(e.g. Leh\'ar et al. 1998).

Rather than discuss well known applications of gravitational lenses
such as determinations of $H_0$ (e.g. Barkana 1998, Impey et al. 1998), 
the cosmological model (Kochanek 1996, Falco et al. 1998),
or the mass distribution of the lenses (e.g. Keeton et al. 1997), we will discuss three
new applications.  First, we will study the fundamental plane of lens
galaxies and its evolution, along with a few preliminary observations 
about the luminosity function of lens galaxies. Second, we will study
extinction and extinction laws in lens galaxies.  Third, and finally,
we will quickly survey the properties of the host galaxies of the
source quasars and AGN.

\begin{figure} % fig.1
\centerline{\psfig{file=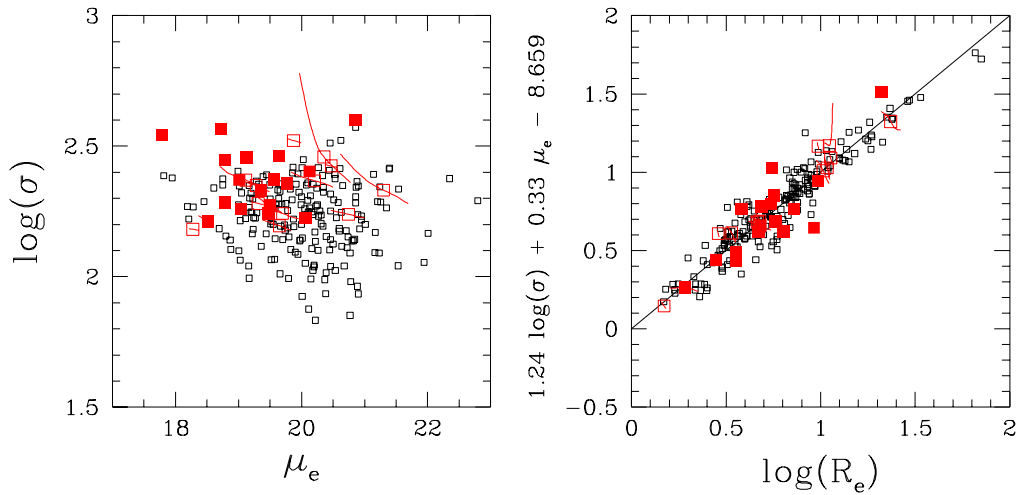,height=2.8in}}
\vspace{5pt}
\caption{
  The Fundamental Plane of lens galaxies.  The right panel shows an
  edge-on view of the FP, and the left panel shows the distribution
  of galaxies in surface brightness $\mu_e$ and velocity dispersion
  $\sigma_c$.  The small open squares are the galaxies from the
  local sample of Jorgensen et al. (1996).  The large filled (open)
  squares are the lenses with known (unknown) lens redshifts.  The
  curves through the open squares show the parameter uncertainties
  created by the remaining redshift uncertainty.}
\end{figure}

\section*{2. The Fundamental Plane of Lenses}

Djorgovski \& Davis (1987) and Dressler et al. (1987) discovered
that early type galaxies show a tight correlation between the effective
radius $R_e$, mean surface brightness $\mu_e$ and central velocity 
dispersion $\sigma_c$ which is now known as the fundamental plane (FP).
The FP is clearly related to the virial theorem, and the differences
between the FP and the relation expected from the virial theorem
are usually interpreted as variations in the mass-to-light ratio
with luminosity.  Van Dokkum et al. (1998) and Pahre (1998) 
 have used the fundamental plane of early-type galaxies
in rich clusters out to $z\sim1$ to study the evolution of the mass-to-light
ratio of the early-type galaxies and to demonstrate that they follow the
predictions for passively evolving stellar populations formed at $z_f\gtorder2$.
There is no comparable study of early-type galaxies in low density 
environments due to observational limitations, although some models
of galaxy formation (e.g. Kauffmann \& Charlot 1998) would predict significantly
different star formation histories for field early-types.  

Most lens galaxies are predicted (e.g. Fukugita \& Turner 1991) and observed 
(e.g. Keeton et al. 1998) to be early-type (E and S0) galaxies in low density
environments.  Many lenses are in groups and very few are in poor clusters.  
To use the FP we need not only $R_e$ and $\mu_e$, which we can obtain from
the CASTLES photometry, but also the central velocity dispersion.  While
we know the mass enclosed by the lensed images with extraordinary accuracy
compared to that obtainable from stellar dynamical studies of even nearby galaxies, 
we have few direct measurements of the central velocity dispersion. We can, however, 
indirectly estimate $\sigma_c$ from the separation of the images 
$\Delta\theta$. Both lens models (e.g. Kochanek 1995) and modern stellar dynamical 
models (e.g. Rix et al. 1997) of early-type
galaxies favor an overall mass distribution corresponding to a flat
rotation curve.  In these models, the image separation depends only on
the dispersion of the dark matter $\Delta\theta \propto \sigma_{DM}^2$,
and stellar dynamical models of local early-type galaxies using the
same mass model show that 
$\sigma_{DM} = \sigma_c$ to remarkable accuracy (Kochanek 1994).  Thus
we estimate the velocity dispersion by 
$ \sigma_c = 225 \left[ (\Delta\theta / 2\farcs91 )(  D_{OS} / D_{LS}) \right]^{1/2}$
km~s$^{-1}$
where $D_{LS}$ and $D_{OS}$ are the lens-source and observer-source distances.

\begin{figure} % fig.1
\centerline{ \psfig{file=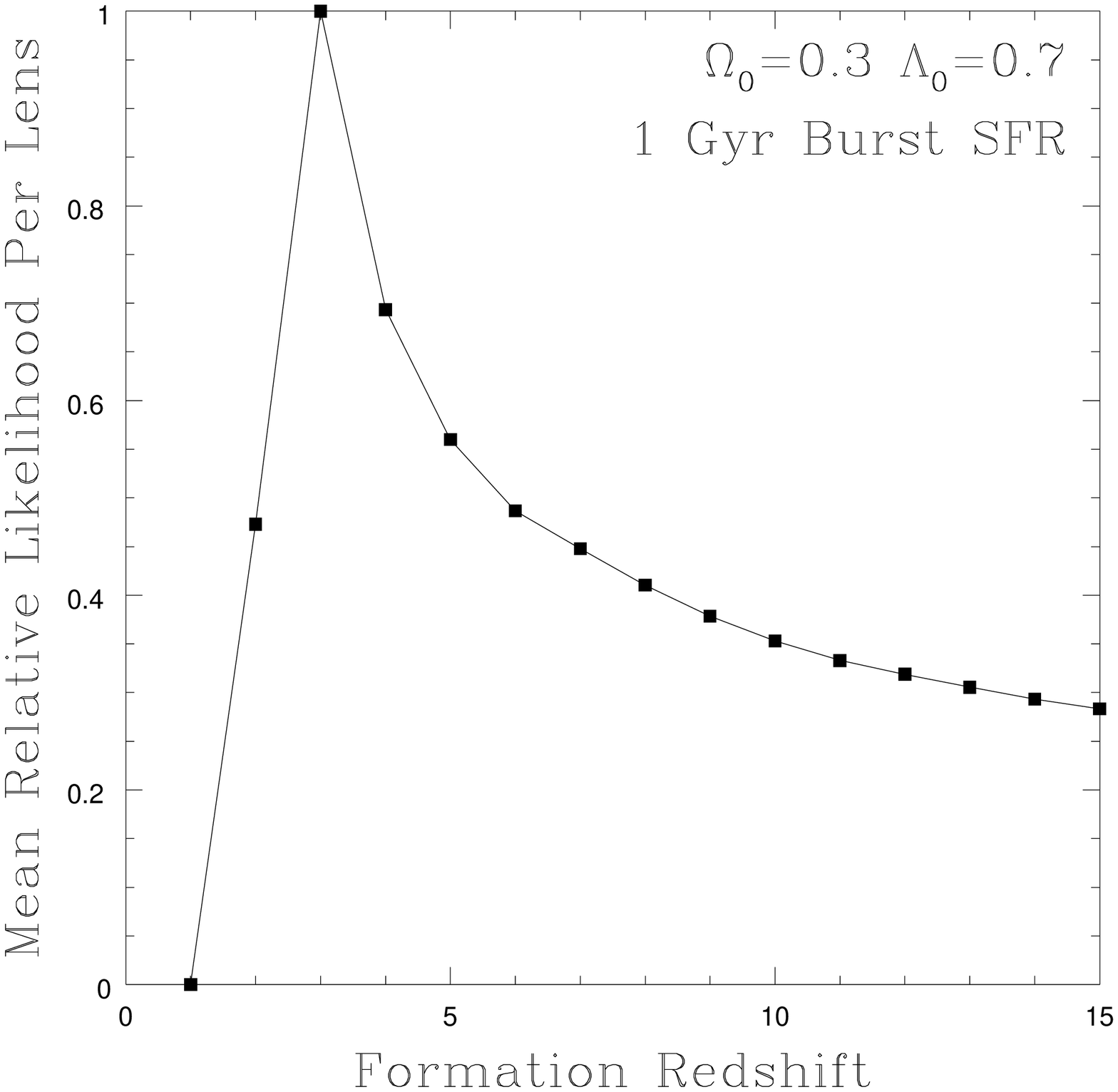,height=3.0in}
             \psfig{file=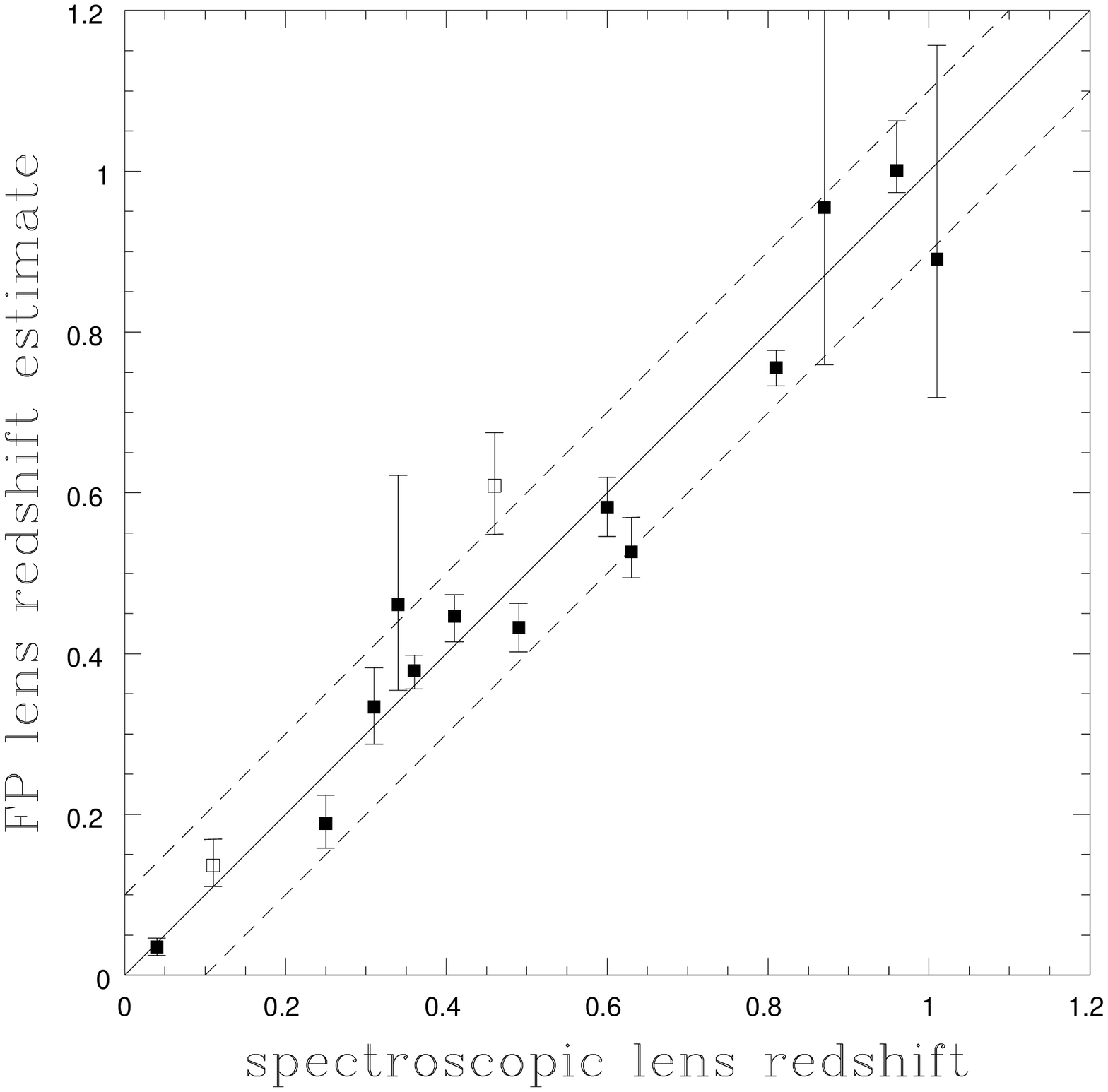,height=3.0in}}
\vspace{5pt}
\caption{
 The relative probabilities of different star formation epochs $z_f$
 for $\Omega_0=0.3$, $\Lambda_0=0.7$, $H_0=65$ km~s$^{-1}$~Mpc$^{-1}$
 and a 1~Gyr burst star formation history.
}
\caption{
  Spectroscopic redshifts versus redshifts estimated from the FP.  Filled
  (open) squares are lenses with known (unknown) source redshifts.
  HST~14113+5211, the most discrepant case, is the open square at
  $z_l=0.46$.
}
\centerline{ \psfig{file=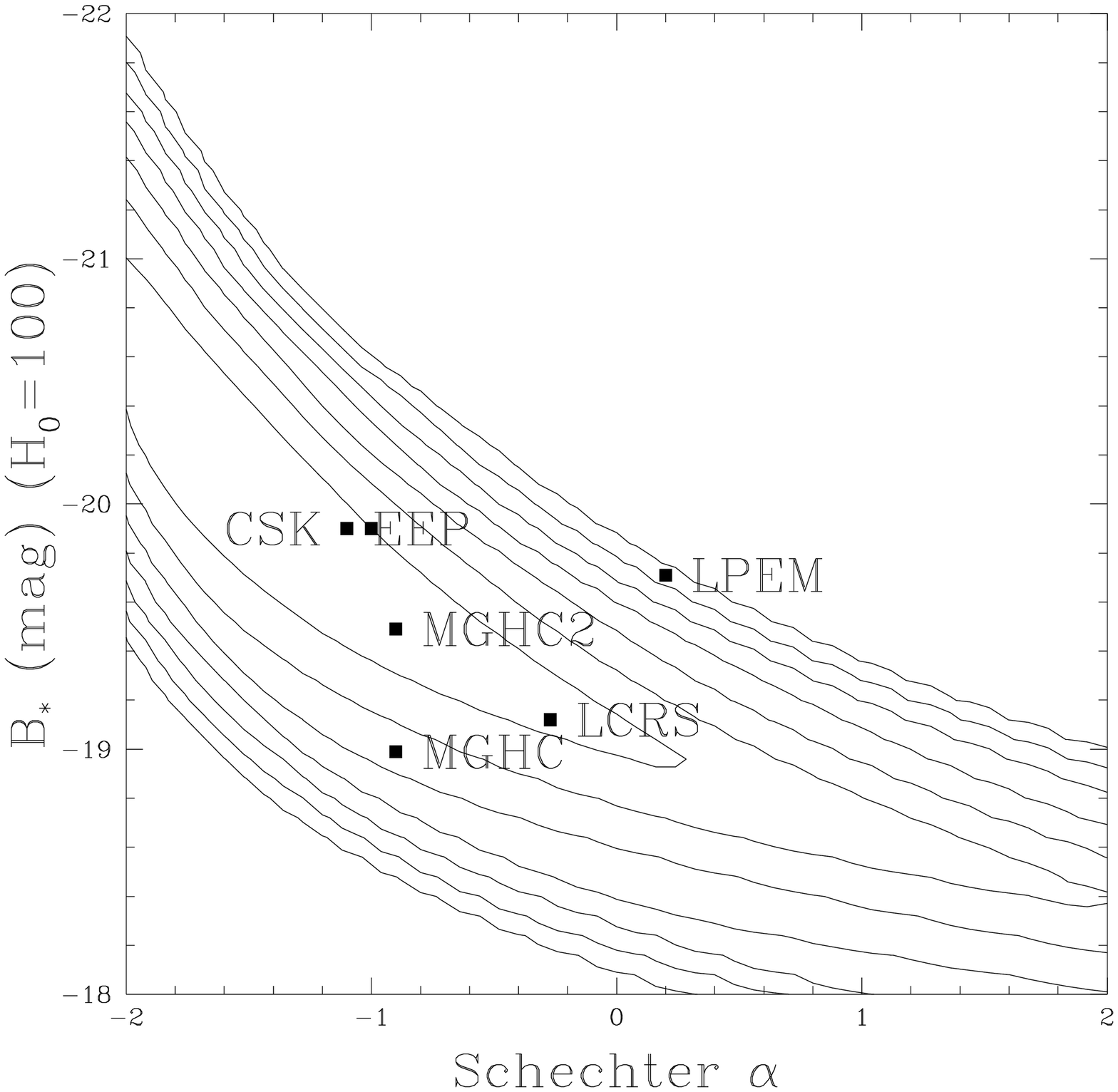,height=3.0in}
             \psfig{file=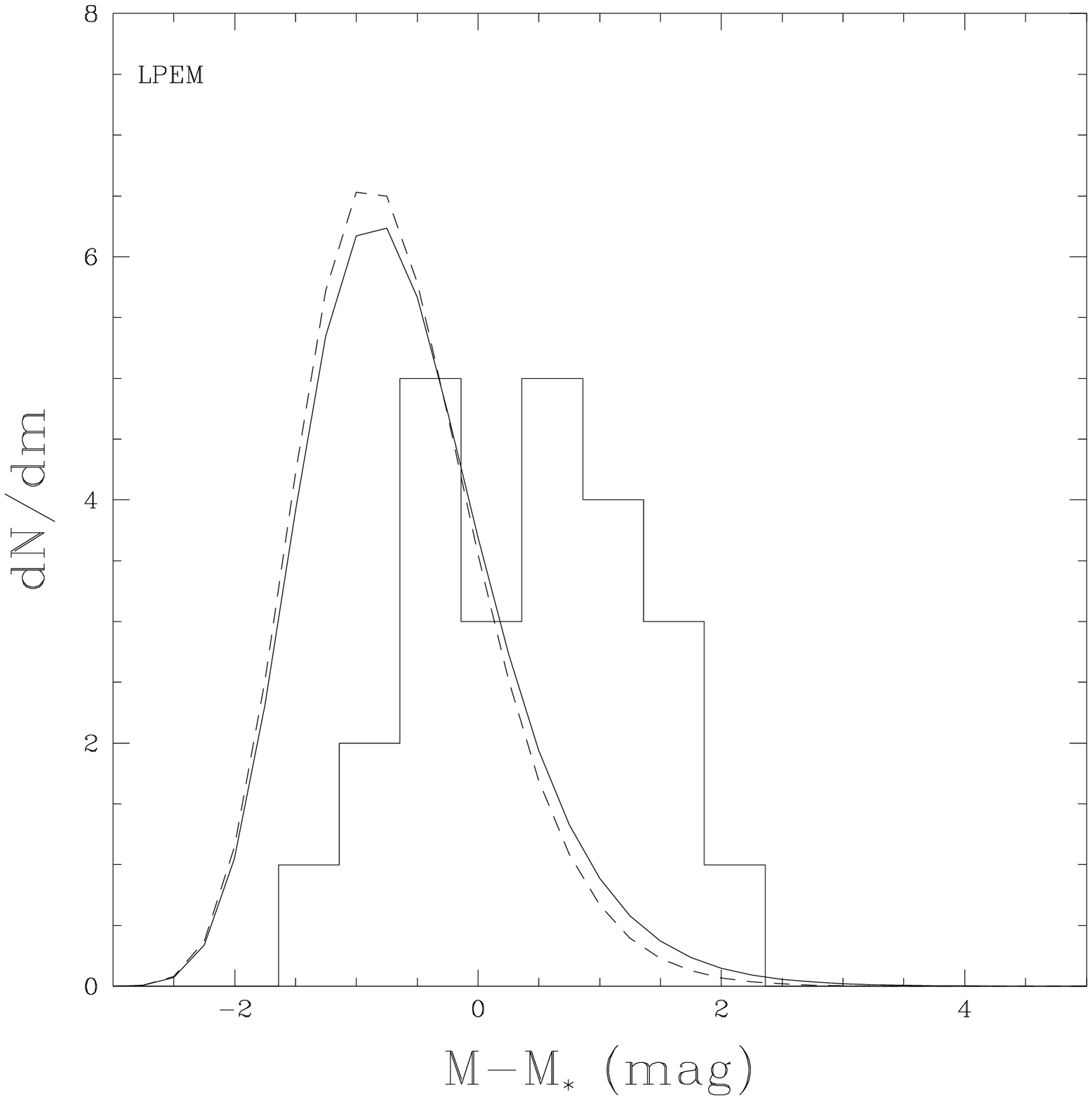,height=3.0in}}
\vspace{5pt}
\caption{
 Left: Likelihood contours spaced by a factor of 10 for
  the Schechter function parameters $\alpha$ and $B_*$.  The points show
 the standard model of Kochanek (1996, labeled CSK) and the five LF models used by
 Chiba \& Yoshii (1998, labeled EEP, MGHC, MGHC2, LCRS, and LPEM).
 Chiba \& Yoshii (1998) used the two most discrepant points
 (LPEM and MGHC) to revise the cosmological limits.
 Right:  An illustration of why the LPEM model fails.
  The observed (histogram) and LPEM predicted (curves) luminosity
  functions of the lenses for an $\Omega_0=0.3$ flat cosmology.
  The dashed (solid) curves include (exclude) a model for the
  selection effects due to finite angular resolution in lens surveys.
}
\end{figure}

We compare the lenses to the local FP by mapping the properties
of the lenses to the redshift of Coma, and then comparing the predicted 
properties of the lenses to the sample
of early-type galaxies in nearby clusters studied by Jorgensen et al. (1996).
To make the transformation we must select a cosmological model (we will use
the currently popular $\Omega_0=0.3$, $\Lambda_0=0.7$ model) and a stellar 
population evolution model.  We used burst models in which star formation
starts at a formation redshift $z_f$ and lasts for 1~Gyr, which provides
a good match to the colors of the lens galaxies for $z_f \gtorder 2$.

Figure 1 shows the FP at Coma and the predicted properties of the lens 
galaxies. {\it With no adjustable parameters other than the
star formation epoch, the vast majority of lenses with known 
redshifts lie on the fundamental plane with little more scatter than
seen in the local samples!}  The success of the comparison again
emphasizes that most lens galaxies are early-type galaxies, and that the
dark matter model we have used to estimate the central velocity dispersion
is making accurate dynamical predictions. Changes in the cosmological model 
tend to make the galaxies slide along the FP, while changes in the star
formation history tend to move the galaxies perpendicular to the FP.
Figure 2 shows the relative probabilities
of the star formation onset redshift $z_f$ for the burst models.
For this particular model the preferred epoch
is $z_f\simeq 3$, rather similar to the results for early-type galaxies
in rich clusters despite the vast difference in environmental densities.

Even though the lens galaxies fall on the present day FP, they are not a
random subset of it.  As the left panel of Figure 1 shows, the lenses are
concentrated toward
high velocity dispersion. We see little difference in the effective radius
and surface brightness distribution of the lens galaxies and the Jorgensen
et al. (1996) local sample.  Note, however, the selection effect
that the lenses with spectroscopic
redshifts tend to have higher surface brightnesses than the lenses with only
estimated redshifts.  The concentration of the lens galaxies at high velocity
dispersions compared to local samples is expected from the strong velocity
dispersion dependence of the probability that a galaxy will be a lens
($\propto \sigma_{DM}^4 \propto \sigma_c^4$).

Many redshifts
for optically selected lenses remain unmeasured because the quasar/galaxy
contrast makes the observations technically challenging, and the 
continuing redshift incompleteness is a severe limitation on using the
lenses to determine the cosmological model.  We can use the constraint
that a lens lies on the FP as a means of estimating unmeasured lens 
redshifts because the predicted physical properties of the lens at Coma
as a function of lens redshift follow a trajectory that is largely
perpendicular to the FP.  Thus, it is only at or near the true lens 
redshift that the galaxy properties will lie on the FP.  Figure 3 compares the 
spectroscopic and FP estimates for the redshifts of 15 lenses.
The rms redshift difference is only $0.06$.  The least accurate estimate
is for HST~14113+5211 (Fischer et al. 1998), where the lens galaxy is in a cluster.
The cluster potential boosts the image separation, leading to an
overestimated galaxy velocity dispersion.  Sometimes the trajectory for
a particular filter moves along the FP for some redshift region, causing
the larger uncertainties seen for two lenses near $z=1$ with
only H band data.  Multicolor data breaks the degeneracy.

Finally, if we know the redshifts of the lens galaxies we can also estimate
the mean luminosity function (LF) of the lenses and compare it to the 
predictions from local estimates of the LF.  The LF of the lens galaxies 
differs from the LF of all galaxies because the lens cross section
rises with luminosity, and we must include the appropriate cross section weighting 
of the LF when we make comparisons.
As emphasized by Kochanek (1996) and Falco et al. (1998), the uncertainties
in the luminosity function of galaxies by type contributes as much to
the uncertainties in the cosmological limits derived from lens statistics
as the Poisson errors arising from the small size of the samples. 
When parametrized by a Schechter function,
$dn/dL=(n_*/L_*)(L/L_*)^\alpha \exp(-L/L_*)$, different local surveys
(e.g. EEP (Efstathiou et al. 1988), LPEM (Loveday et al. 1992), MGHC (Marzke et al. 1994),
 LCRS (Lin et al. 1996))
find mutually discrepant values for the faint end slope $\alpha$, break
luminosity $L_*$, and number density of early-type galaxies.  

Figure 4 shows the likelihood of fitting the observed LF of the lens 
galaxies and the separation distribution of the images as a function of the
Schechter function parameters $\alpha$ and $B_*$ (the absolute B magnitude 
corresponding to $L_*$) in an $\Omega_0=0.3$, $\Lambda_0=0.7$ cosmology.
We mark the central point of the standard model used by Kochanek (1996) and
Falco et al. (1998) and five alternative models used by Chiba \& Yoshii
(1998).  The value of $\alpha$ is that from the original LF surveys (listed above),
while the value of $B_*$ is an estimate by Chiba \& Yoshii (1998) after converting
from the photometric band of the original survey.  The two models selected by 
Chiba \& Yoshii (1998) to revise the cosmological limits (LPEM and MGHC)
are the two models most discrepant with the
lens data, probably because of problems in the estimate of $B_*$.  
Figure 4 reveals why the LPEM model fails.  
The likelihood contours for the lens data show a degeneracy 
between $B_*$ and $\alpha$ which is very similar to the apparent
degeneracy that links most of the local LF estimates.
These LF comparisons are extremely preliminary, but with the full
sample it should be possible to constrain $B_*$ and $\alpha$ more
accurately, while measuring the changes in the comoving density
$n_*$ with redshift.

\section*{3. Extinction}

We possess little direct information on extinction in galaxies outside
the Local Group and almost none on early-type galaxies (see reviews by
Mathis 1990, Fitzpatrick 1998).  Accurate
extinction estimates almost always depend on knowing the intrinsic
spectrum of the reddened object, which is generally true only of stars.
Once inferences about extinction depend on modeling the fluxes of
stellar populations mixed with dust, the accuracy drops dramatically
(e.g. Witt et al. 1992).
Extinction laws are measured almost exclusively in the Galaxy, the LMC
and the SMC.  No accurate extinction curve is measured in an early-type
galaxy, although several studies (e.g. Warren-Smith \& Berry 1983) suggest 
that dust in early-type
galaxies may be quite different from ``standard'' Galactic dust.
Moreover, since both the mean metallicity and star formation rates are
strong functions of redshift, it would be surprising if the mean
extinction curve failed to evolve with redshift.

We can use the lenses to determine the differential extinction between
the lensed images from the variation in the flux ratios with wavelength.
If there is sufficient dust and wavelength coverage, the extinction law
can be determined (Nadeau et al. 1991), and it may be possible to determine
the redshift of the dust (Jean \& Surdej 1998).  
The magnitude difference between two images $i$ and $j$ as
a function of wavelength $\lambda$ is
\begin{equation}
  m_i(\lambda)-m_j(\lambda) = -2.5 \log \left( M_i / M_j \right)
     + (E_i -E_j) R \left( \lambda/(1+z_d) \right)
\end{equation}
where $M_i/M_j$ is the magnification ratio,
$E_i-E_j$ is the extinction difference ($\Delta E(B-V)$), and $R(\lambda/(1+z_d))$
is the extinction law in the rest frame of the dust.  Systematic
errors arise if the magnification ratio depends on wavelength 
or temporal variations by the source mimic a wavelength dependence.

\begin{figure} % fig.1
\centerline{\psfig{file=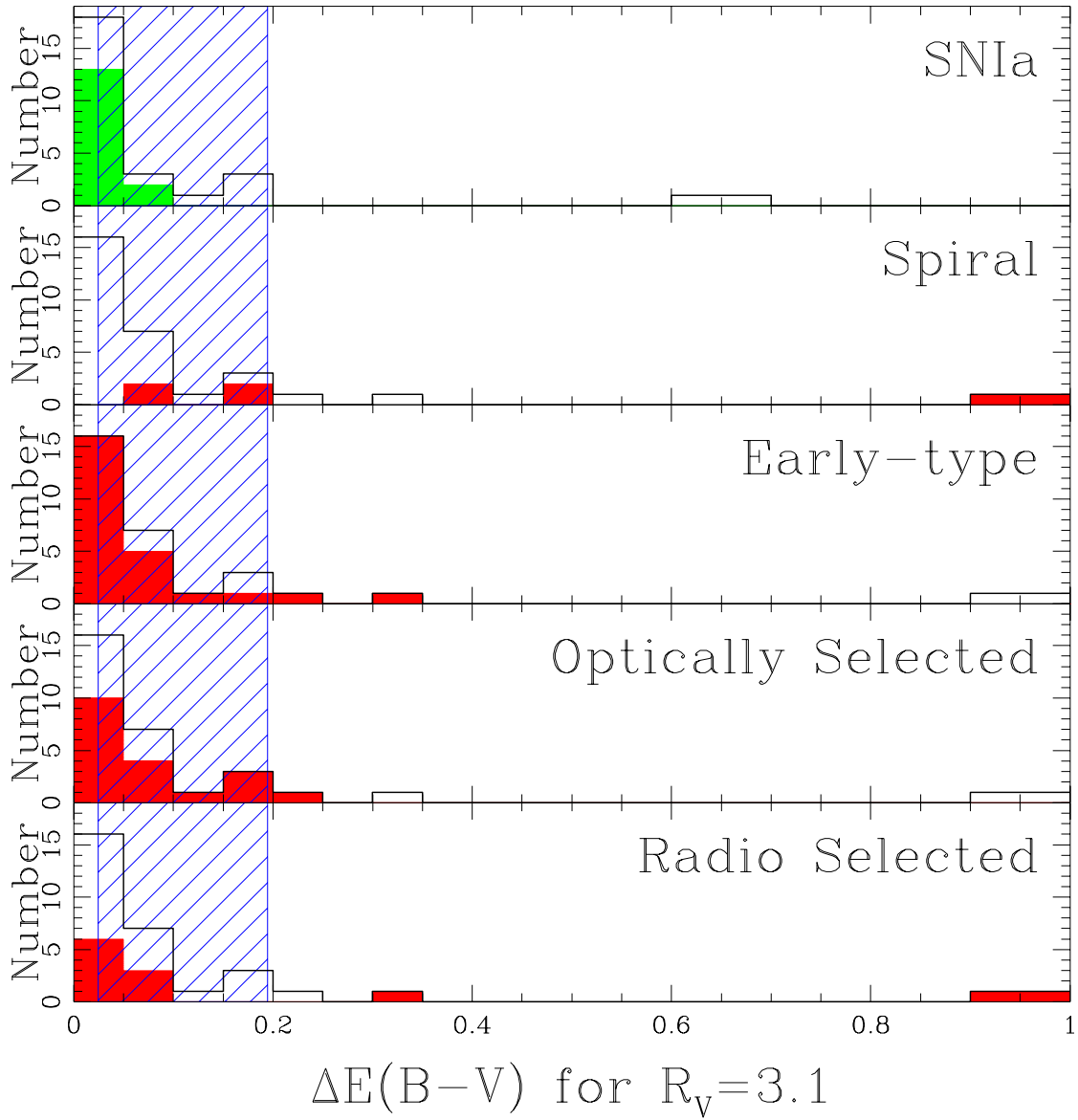,height=6.5in}}
\vspace{5pt}
\caption{
 Histograms of differential extinctions.  The solid histogram is the
 total sample, and the shaded histogram shows the distribution of
 the (from bottom to top) radio-selected, optically-selected, early-type and late-type
 subsamples.  The hatched region shows the mean extinction estimated
  from a comparison of the statistics of lensed quasars and radio
 sources by Falco et al. (1998).  The two objects in the high extinction
  bin actually have $\Delta E=1.0$ and $3.0$ mag.  The top panel shows
  the extinction distribution of Type Ia supernovae from Riess et al. (1998)
  for low redshift (open) and high redshift (shaded).  Note, however, that
  11 of the 15 high redshift supernovae have negative extinctions which
  have been reset to zero.  }
\end{figure}

Figure 5 presents a histogram of differential extinctions on 37 lines of
sight in 23 lens galaxies.  The median rest frame differential extinction 
of the optically selected lenses is $\Delta E(B-V)=0.04$ mag and the median
for the radio selected lenses is $0.07$ mag.  The distributions for optical 
and radio selected lenses are similar except for the two radio-selected
lenses with high differential extinctions.  Both B~0218+357 and PKS~1830--211
are face-on spiral galaxies (Leh\'ar et al. 1998) with high molecular gas content (e.g.
Wiklind \& Combes 1995, 1996).  There is no correlation of the differential
extinction with impact parameter, which suggests that the diffuse dust is patchy.  
For comparison to the differential extinction, we had previously obtained
an estimate of the mean extinction in lens galaxies of $A_B=0.58\pm0.45$
mag by comparing the statistics of radio and optically selected lenses
(Falco et al. 1998).  Thus the mean extinction is comparable to the
differential extinction, also consistent with a patchy dust distribution.

\begin{figure} % fig.1
\centerline{\psfig{file=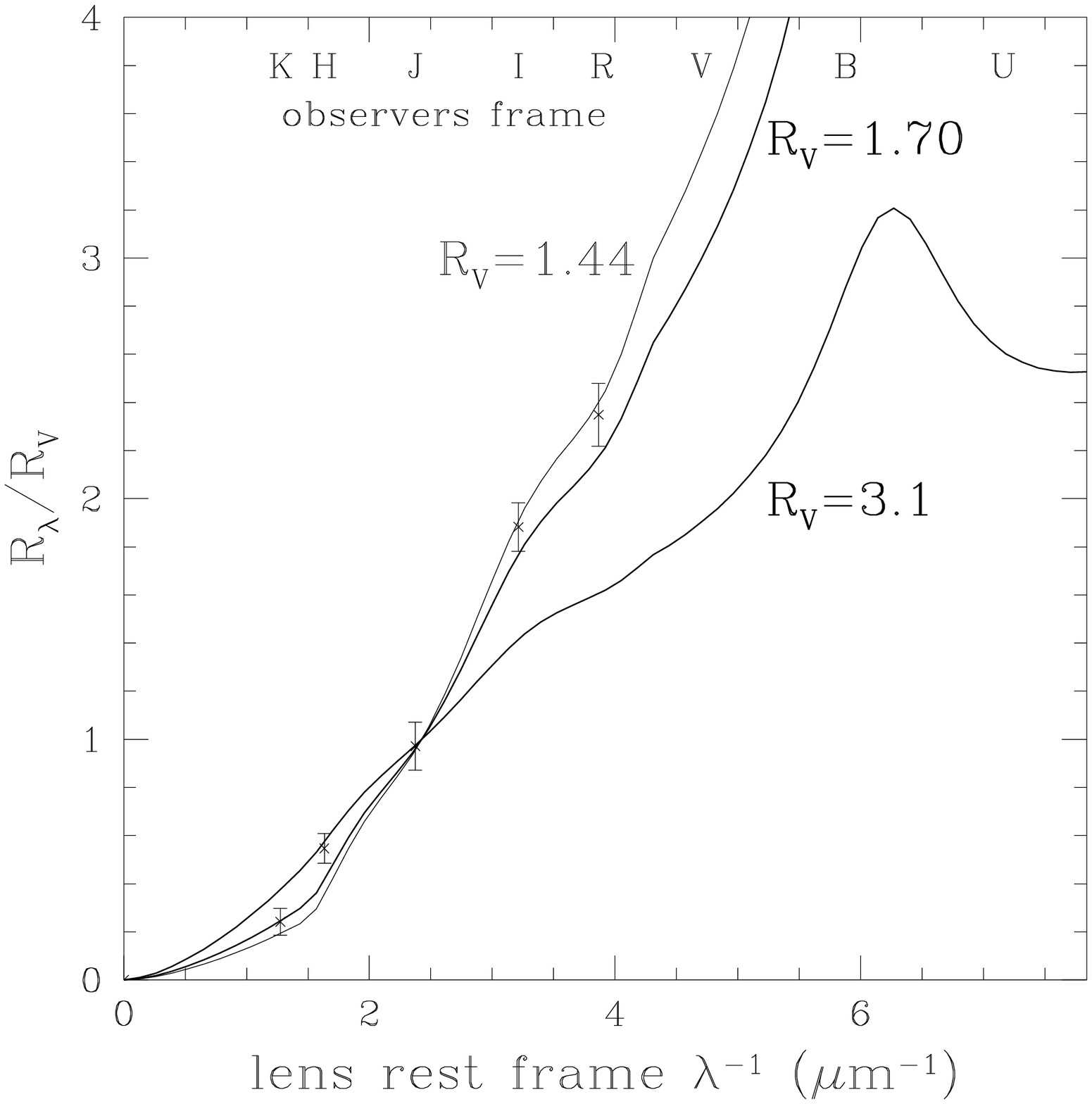,height=3.0in}
            \psfig{file=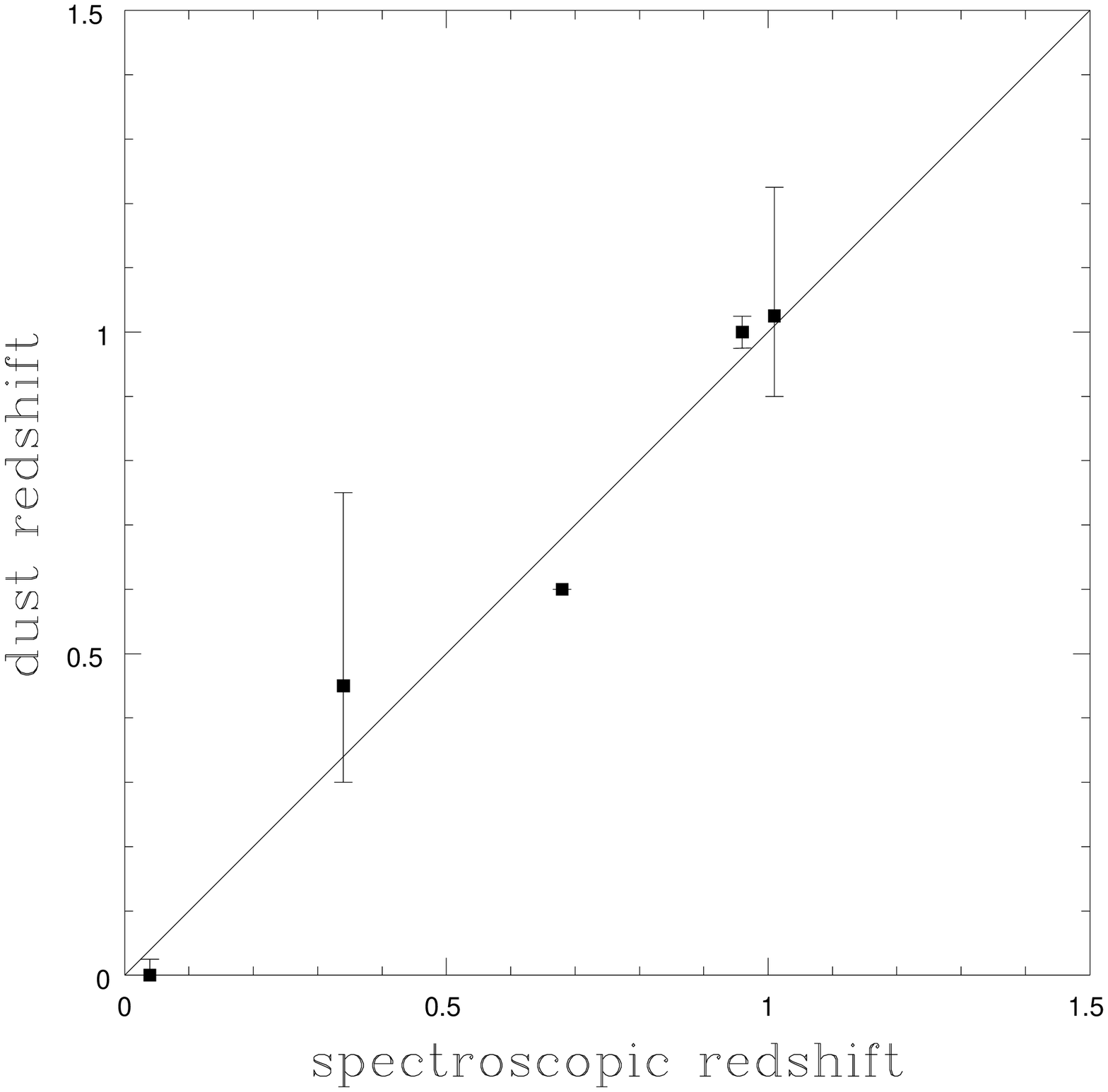,height=3.0in}}
\vspace{5pt}
\caption{
  The extinction law in the $z_l=0.96$ elliptical lens galaxy 
  MG~0414+0534.  The non-parametric extinction curve is shown by
  the points along with the Cardelli et al. (1989) model with the
  same $R_V=1.44\pm0.09$.  The best Cardelli et al. (1989) parametric
  model has $R_V=1.7\pm0.1$.  The standard $R_V=3.1$ Galactic curve
  is shown for comparison.
  }
\caption{
  Spectroscopic versus dust redshifts for five lens systems.  In
  order of increasing redshift they are Q~2237+0305, B~1422+231,
  B~0218+357, MG~0414+0534,
  and MG~2016+112.  Using $\Delta\chi^2=1$ appears to underestimate the 
  uncertainties in the dust redshifts.  The mean error is
  $\langle z_{dust} - z_l \rangle = 0.01 \pm 0.07$. 
}
\end{figure}

\begin{figure} % fig.1
\centerline{\psfig{file=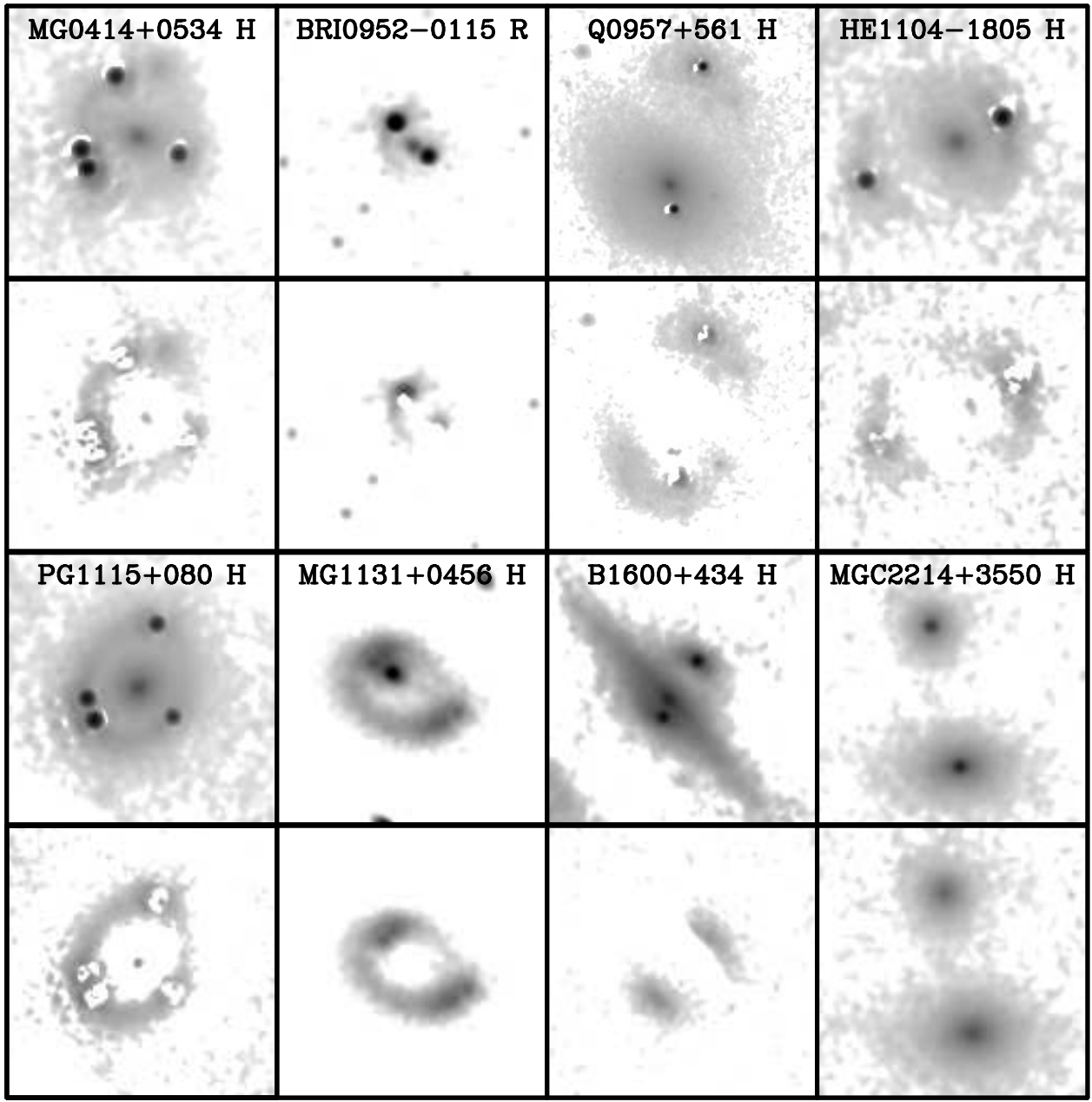,height=7.0in}}
\vspace{5pt}
\caption{
  Host galaxies.  The top image shows the full image, and the bottom image shows
  the host after subtracting the lens and the active nucleus.  All images are
  5\farcs8 square except for Q~0957+561 which is 11\farcs5 square.}
\end{figure}

For systems with sufficient extinction and wavelength coverage we
can estimate the extinction law.  Figure 6 shows the extinction law
of the $z_l=0.96$ elliptical lens in MG~0414+0534 
derived both parametrically using the Cardelli et al. (1989) models
and non-parametrically.  The values of $R_V=1.44\pm0.09$ in the
non-parametric models and $R_V=1.7\pm0.1$ in the parametric models
are well below the standard Galactic value of $R_V=3.1$.  In fact,
standard
Galactic dust is ruled out at a $\Delta\chi^2=56$, and we confirm
the local indications that dust in early-type galaxies may have a 
different mean extinction law than that of the Galaxy.  

Following Jean \& Surdej (1998) we can also estimate a dust redshift
for the systems using the dependence of the extinction at observed
wavelength $\lambda$ on the extinction curve at the rest wavelength
in the dust $R(\lambda/(1+z_d)$).   The determination of dust redshifts
requires better data than determining the extinction curve because
it depends on detecting the deviations of the extinction curve from
a self-similar power law $R_\lambda \propto \lambda^{1.7\pm0.1}$ (e.g.
Mathis 1990).  For
five cases we can compare the spectroscopic lens redshift to the dust
redshift (see Figure 7), including MG~0414+0534. The agreement
is remarkably good. Moreover, the fact that the anomalous extinction 
law predicts the observed redshift of the MG~0414+0534 lens strongly
suggests that the result is not a consequence of systematic errors.

We are measuring extinctions and extinction laws at redshifts and 
impact parameters very similar to the those of the Type Ia supernovae used by 
Perlmutter et al. (1997) and Riess et al. (1998) to determine the
cosmological model.  It is critical to the cosmological determination
that the supernova fluxes be accurately corrected for extinction.
We can make four observations about the supernovae from the lens samples.
First, if even early-type galaxies at these redshifts contain dust, it
would be surprising if the supernova samples showed no dust.  
Second, from the radio lenses like MG~0414+0534, we can rule out the 
existence of dust which is gray in the optical.  Where we can measure the 
extinction curves in lens galaxies they resemble the Cardelli et al. 
(1989) parametrized forms.  Third, galaxies at high redshift, like galaxies 
at low redshift, show a range of extinction curves.  It is dangerous 
to assume that the extinction curve will match a mythical standard extinction law.
Fourth, we could not compare the lens and supernovae extinction distributions.
Perlmutter et al. (1997) do not estimate extinctions.  While Riess et al. (1998)
estimate extinctions, the distribution is peculiar because 11 of 15
supernovae have negative estimated extinctions.  Simple 
statistical tests show that the preponderance of negative extinctions
in the sample is inconsistent with the assumption
that they are produced by random photometric errors with the stated
uncertainties at a slightly greater than 2-$\sigma$ confidence level. 
  
\section*{4. A Quick Tour of Host Galaxies}

A large fraction of the lens systems now show arc and ring images
of the quasar or AGN host galaxies either in the optical or in
the infrared.  Figure 8 shows a sample of the hosts.  Seven of
the eight hosts are new discoveries.  
  MG~0414+0534 ($z_s=2.64$) is the system used
    to determine the extinction law in \S3.  
  BRI~0952--0115 ($z_s=4.5$)
    is the highest redshift detection of a host galaxy.
  Q~0957+561 ($z_s=1.41$) shows two enormous arc images of the host whose
    morphologies essentially rule out the popular Grogin \& Narayan (1996) 
    models for the system.  These models predict a radially stretched image 
    of the Northern host relative to the lens galaxy, while we observe a tangentially stretched image.
  HE~1104--1805 ($z_s=2.32$) shows two arc images of the host.
  PG~1115+080 ($z_s=1.72$) has an Einstein ring image of the host galaxy which 
    could be used to break the degeneracies in the estimates of $H_0$ for the 
    system if NICMOS is repaired (Impey et al. 1998). 
  B~1600+434 ($z_s=1.59$) shows two images of the host straddling the bulge of the lens.
  MGC~2214+3550 ($z_s=0.88$) is a beautiful example of a binary quasar.  The
    lower source is a radio loud quasar and the upper is not, with
    a radio flux ratio of $> 80$.  Both H band sources are perfectly
    modeled by a point source for the quasar at the center of a
    de Vaucouleurs profile host.
  MG~1131+0456 ($z_s=?$) has a spectacular infrared ring image of the host
   (Kochanek et al. 1998).  The ring is 4--5 times brighter than the
   $z_l\simeq0.85$ lens galaxy in the
   infrared and virtually invisible in the optical.  The red
   colors of the system are due to the flux from the stars in
   the host galaxy rather than to dust in the lens.

We have started to estimate the properties of the host galaxies,
and the two generic statements seem to be that many correspond
to sub-$L_*$ galaxies for their redshifts and that they are 
relatively blue.  In many cases the host galaxies are actually
bluer in their I--H colors than the early-type lens galaxies.
While the very luminous hosts of radio-loud objects (e.g. 
      MG~1131+0456) have been seen previously, our lensing results 
      constitute the first secure detection for a sample of
      radio-quiet hosts. While in nearby quasars the host luminosities
      do not depend on the radio properties, we find that at $z>1$, the
      radio loud hosts are on average 2 mag brighter than
      the radio quiet ones. One possible interpretation is that
      the radio flux as well as the rest-optical luminosity get 
      boosted during star-burst phases.

\section*{5. Summary}

The next ten years will be the period when gravitational lenses make their most
dramatic scientific impact.  As recently as two years ago, there were too few
lenses to attack many of the most interesting scientific problems,
and ten years from now there will be several hundred lenses and progress
will again slow. Today, with 40--50 lens systems and 4--5 time delay
determinations  we are at the cusp where science using gravitational lenses
will advance most rapidly.  The CASTLES project in combination with archival 
HST observations of gravitational lenses now has about half of the final data
set for the 47 currently known lenses, but new lenses are being discovered
almost as fast as HST is observing the old lenses.

We can use this explosion in data to dramatically expand the range of 
scientific problems that gravitational lenses can attack.  In this
short review we have illustrated only three new examples of gravitational 
lenses as tools.  The fundamental plane of lens galaxies shows that most
lenses are normal early-type galaxies, that early-type galaxies in low density
environments are very similar to those in the centers of rich clusters, and
that the stellar populations of the early-type lenses must have formed at
$z_f \gtorder 2$.  The differential extinction in lens galaxies shows that
early-type galaxies contain modest amounts of diffuse dust.  The amount of
dust is sufficient to bias cosmological limits based on the statistics 
of lensed quasars, as already discovered by Falco et al. (1997) in their
comparison of the statistics of lensed quasars and radio sources.  The dust
in the lenses can be used to determine both extinction laws and lens redshifts.
Finally, HST images of lenses, particularly infrared images, commonly show
arc and ring images of the quasar or AGN host galaxies.  The lens magnification
pulls the host galaxy out from under the bright central point source and makes
it significantly easier to detect and model the properties of the host galaxies.

\medskip
\noindent Acknowledgements: Support for the CASTLES project was provided by NASA through grant numbers
GO-7495 and GO-7887 from the Space Telescope Science Institute, which is operated
by the Association of Universities for Research in Astronomy, Inc.
CSK was also supported by the NASA Astrophysics Theory Program grant NAG5-4062.  
HWR is also supported by the Alfred P. Sloan Foundation.

\end{document}